\documentclass[]{interact}

\usepackage{epstopdf}

\usepackage[numbers,sort&compress]{natbib}
\bibpunct[, ]{[}{]}{,}{n}{,}{,}
\usepackage{appendix}
\usepackage{fancyhdr}
\usepackage{float}
\usepackage{setspace}
\usepackage{hyperref} 

\usepackage{booktabs}
\usepackage{multirow}
\usepackage{algorithm}
\usepackage[noend]{algpseudocode}

\usepackage{mathtools}
\usepackage{amsmath, amssymb, bbm}

\usepackage{graphicx}
\usepackage{subcaption}
\usepackage{tikz}
\usetikzlibrary{positioning}

\usepackage{epsfig} 

\theoremstyle{plain}

\theoremstyle{definition}

\begin{document}

\articletype{ARTICLE TEMPLATE}

\title{Counterfactual Q-Learning via the Linear Buckley–James Method for Longitudinal Survival Data}

\author{
\name{Jeongjin Lee\textsuperscript{a} and Jong-Min Kim\textsuperscript{b}\thanks{CONTACT Jong-Min Kim. Email: jongmink@morris.umn.edu}}
\affil{\textsuperscript{a}Division of Biostatistics, The Ohio State University, 281 W Lane Ave, Columbus, OH 43210, U.S.A.; 
\textsuperscript{b}Statistics Discipline, Division of Science and Mathematics,
University of Minnesota-Morris, Morris, MN 56267, USA \& EGADE
Business School, Tecnol\'ogico de Monterrey, Ave. Rufino Tamayo, Monterrey 66269, Mexico.}
}

\maketitle

\begin{abstract}
Treatment strategies are critical in healthcare, particularly when outcomes are subject to censoring. This study introduces the Counterfactual Buckley-James Q-Learning framework, which integrates the Buckley-James method with reinforcement learning to address challenges posed by censored survival data. The Buckley-James method imputes censored survival times via conditional expectations based on observed data, offering a robust mechanism for handling incomplete outcomes. By incorporating these imputed values into a counterfactual Q-learning framework, the proposed method enables the estimation and comparison of potential outcomes under different treatment strategies. This facilitates the identification of optimal dynamic treatment regimes that maximize expected survival time. Through extensive simulation studies, the method demonstrates robust performance across various sample sizes and censoring scenarios, including right censoring and missing at random (MAR). Application to real-world clinical trial data further highlights the utility of this approach in informing personalized treatment decisions, providing an interpretable and reliable tool for optimizing survival outcomes in complex clinical settings.
\end{abstract}


\section{Introduction}
\label{s:intro}

\sloppy

Personalized treatment strategies have improved modern healthcare by tailoring interventions to individual patient needs. Dynamic treatment regimes (DTRs) \citep{moodie2012q, goldberg12, wahed2013evaluating, huang2014optimization, moodie2014q, song2015penalized, wallace2015doubly, kosorok2015adaptive, xu2016bayesian, simoneau2020estimating, ly23, li2024dynamic} provide a systematic framework for making sequential treatment decisions based on patients’ evolving health conditions. This approach is particularly beneficial for addressing censored data, where survival times may be incomplete due to patient dropout or study termination.

Survival analysis \citep{kleinbaum1996survival, moore2016applied}, which investigates the timing of specific events, is a fundamental aspect of medical research. Despite its importance, the frequent occurrence of censoring complicates both the analysis and interpretation of survival data. A commonly used approach to manage censoring is the Cox proportional hazards model \citep{cox1972regression}, which, while effective in many scenarios, has some drawbacks. These include the reliance on the proportional hazards assumption and the potentially counterintuitive interpretation of hazard ratios (HRs), which reflect comparisons of event rates between groups and are influenced by the baseline hazard function.

The accelerated failure time (AFT) model \citep{bu79, wei1992accelerated, ji03, lambert2004parametric, ji06, ze07} provides an alternative framework with several advantages. Unlike the Cox model, the AFT approach employs a linear regression framework, often using a logarithmic transformation of the response variable, making its results more interpretable and independent of the proportional hazards assumption. The Buckley-James (BJ) method \citep{bu79} is a widely used estimator in the context of the AFT model due to its efficiency and minimal assumptions regarding censoring mechanisms. Extensions of the BJ method, including penalized versions, have been developed to handle high-dimensional censored data effectively, with studies focusing on their theoretical properties in large samples \citep{jo08, wa08, jo09, li14, soret2018lasso, yin2022bayesian, li2021robust, wang2022regularized, lee2024censored}.

Reinforcement learning, particularly Q-Learning \citep{wa92, clifton2020q}, provides a robust framework for optimizing personalized treatment strategies through sequential decision-making. In healthcare, treatment decisions are often made iteratively, adjusting to a patient’s changing health status over time. Q-Learning facilitates the development of dynamic treatment regimes (DTRs) by leveraging cumulative rewards to identify policies that maximize long-term clinical outcomes. This approach enables the personalization of interventions based on patient-specific data collected at each decision point.
Despite its potential, applying Q-Learning in healthcare presents unique challenges, particularly in survival analysis, where outcomes are often censored. Censoring occurs when the event of interest is not observed within the study period, leading to incomplete data. This complicates the estimation of rewards, a critical component of Q-Learning, and can result in biased policy recommendations.

To address these challenges, we propose the Linear Buckley-James Q-Learning framework, which integrates the Buckley-James method \citep{bu79, la91, ji06} to effectively manage censored survival data while optimizing treatment strategies. The Buckley-James method addresses censoring by imputing unobserved survival times through a combination of observed data and the conditional expectation of censored outcomes, offering a robust foundation for analysis. These imputed survival times are incorporated into the Q-Learning framework, enabling iterative refinement of treatment policies based on updated survival information. This systematic integration facilitates the accurate estimation and comparison of treatment effects across different strategies. The Buckley-James method serves as a critical component, enhancing the precision and robustness of the overall approach.

A key strength of our approach lies in its ability to handle censored survival data while enabling meaningful comparisons of imputed survival times across various treatment strategies. While many methods exist for handling censored data, our approach uniquely facilitates direct comparison across different treatment regimes by leveraging imputation techniques that preserve survival time relationships.
For example, Goldberg and Kosorok \citep{goldberg12} developed a dynamic treatment regime estimator using Q-learning, addressing challenges in handling censoring and varying treatment stages through inverse probability weighting. Huang et al. \citep{huang2014optimization} optimized treatment strategies in recurrent disease trials using backward recursion but did not focus on direct survival time comparisons. Simoneau et al. \citep{simoneau2020estimating} extended Q-learning to non-censored outcomes using weighted least squares for time-to-event data, while Cho et al. \citep{cho2023multi} used generalized random survival forests for dynamic treatment regimes. 
Unlike these methods, our framework explicitly accounts for censored observations while ensuring that imputed survival times allow for robust comparisons across treatment strategies under counterfactual framework.
This capability is critical in evaluating treatment efficacy where direct observation of survival times is limited.

The structure of this article is as follows: Section 2 introduces the Linear Buckley-James Q-Learning framework. Section 3 describes the generation of synthetic patient datasets and provides details of the comprehensive simulation study conducted. Section 4 presents the analysis of real-world clinical data. 
Finally, Section 5 discusses the results and broader implications. Sample R code for implementing the proposed methodology is available at \url{https://github.com/Jeongjin95/BJ-Q-Learning}.

\section{Method}

This section introduces the Linear Buckley-James Q-Learning framework, which combines the Buckley-James (BJ) imputation method with reinforcement learning. The framework is specifically designed to handle right-censored survival data while optimizing multi-stage treatment strategies.

\subsection{Preliminary}
Consider a longitudinal clinical trial conducted over \( K \) decision points, indexed by \( k = 1, \ldots, K \), where each stage \( k \) corresponds to the interval between two clinical visits. 
Visits may occur on a fixed schedule (e.g., every 30 days) or be triggered by clinical events (e.g., fever onset, worsening symptoms).
At each stage \( k \), the clinician observes patient-specific, time-varying covariates \( X_{i,k} \) (e.g., blood pressure, biomarker levels, symptom scores), and a treatment \( A_{i,k} \in \mathcal{A} \) is randomly assigned according to a prespecified protocol. The treatment set \( \mathcal{A} \) may include binary options (e.g., drug vs. placebo), ordinal doses (e.g., low, medium, high), or categorical choices (e.g., standard of care, experimental therapy). 
The patient’s observed history at stage \( k \) is denoted by
\[
H_{i,k} = \left(B_{i,0}, X_{i,1}, A_{i,1}, \ldots, X_{i,k}, A_{i,k} \right),
\]
where \( B_{i,0} \) are baseline covariates (e.g., age, sex, genetic markers), and the sequence \( \{X_{i,l}, A_{i,l}\}_{l=1}^k \) captures the patient's evolving clinical status and treatment assignments up to stage \( k \).
Here, \( i = 1, \ldots, n \) indexes the \( n \) patients enrolled in the trial.

We define the observed outcome at stage \( k \) based on the true time-to-event \( T_{i,k} \), which represents the elapsed time between the \((k{-}1)\)th and \(k\)th clinical visits, and the overall censoring time \( C_{i} \), which is the time at which follow-up is lost or the study administratively ends. The observed duration at stage \( k \) is given by:
\[
Y_{i,k} = \min(T_{i,k}, C_{i}),
\]
with the corresponding censoring indicator defined as
\[
\delta_{i,k} = \mathbbm{1}(T_{i,k} \leq C_{i}),
\]
which equals 1 if the visit time or death is observed before censoring, and 0 otherwise. Censoring occurs when the event of interest is not observed due to reasons such as loss to follow-up, patient withdrawal, or administrative termination of the study.

\begin{figure}[h!]
  \centering 
  \includegraphics[scale=0.4]{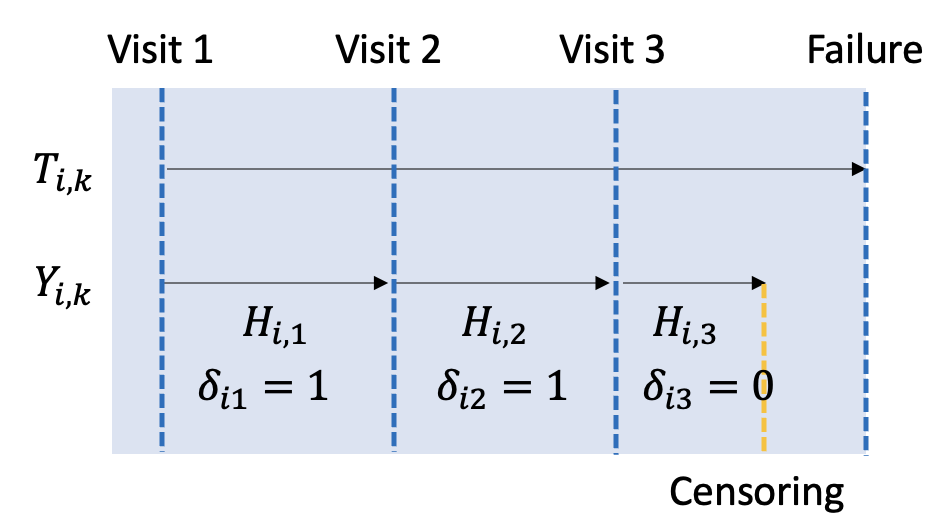} 
  \caption{Detailed Example of a Particular Patient}
  \label{fig:patient}
\end{figure}

Figure~\ref{fig:patient} provides a detailed example of a particular patient’s follow-up over three visits. At each visit, patient information (\( H_{i,k} \)) is collected. The patient’s visit times are observed without censoring at Visits 1 and 2 (\( \delta_{i1} = 1, \delta_{i2} = 1 \)), indicating that follow-up continues beyond these visits. However, at Visit 3, the outcome is censored (\( \delta_{i3} = 0 \)), meaning the failure event of interest was not observed before the study ended. This figure illustrates how longitudinal follow-up data captures both event times and censoring across multiple stages.

Dynamic treatment regimes (DTRs) provide a framework for optimizing sequential treatment decisions based on individual patient characteristics. A DTR consists of a sequence of decision rules \( \{d(H_1), \ldots, d(H_K)\} \), where each rule maps a patient’s history \( H_{i,k} \) to a treatment \( A_{i,k} \). The goal is to adapt treatment choices at each stage to maximize long-term outcomes such as survival, especially in the presence of censoring and time-varying covariates.
The overall survival time \( T_i \) is defined as the sum of the stage-specific survival times, weighted by an entry indicator \( \eta_{i,k} \), which equals 1 if patient \( i \) enters stage \( k \) and 0 otherwise:
\begin{equation}
T_{i,\text{cum}} = \sum_{k=1}^K \eta_{i,k} T_{i,k}.
\end{equation}
The objective of a dynamic treatment regime (DTR) is to maximize the expected cumulative survival time. Formally, the total survival time under a given treatment sequence \( a = \{a_1, \ldots, a_K\} \) is defined as:
\begin{equation}
T_{i,\text{cum}}(a) = \sum_{k=1}^K \eta_{i,k} T_{i,k}(a_k),
\end{equation}
where \( T_{i,k}({a_k}) \) represents the potential survival (or visit) time at stage \( k \) if treatment \( a_k \) is assigned, and \( \eta_{i,k} \) is an indicator that patient \( i \) has survived and reached stage \( k \).
The optimal dynamic treatment regime (DTR) for individual $i$, denoted as \( d_i^{\text{true}} = \{d_{i,1}^{\text{true}}, \ldots, d_{i,K}^{\text{true}}\} \), is the sequence of decision rules that maximizes the expected cumulative survival time across all stages:
\begin{equation}
d_i^{\text{true}} = \arg\max_{(a_1, \ldots, a_K) \in \mathcal{A}^K} \mathbb{E}\left[\sum_{k=1}^K \eta_{i,k} T_{i,k}(a_k) \right],
\end{equation}
where \( \mathcal{A} \) is the set of available treatments.
The corresponding Q-function at stage \( k \) under treatment \( a_k \) is defined as the expected cumulative survival time from stage \( k \) onward, given the treatment and history up to that point:
\begin{equation}
\label{eq:true_q_function}
Q_k^*(H_{i,k}(a_k)) = \mathbb{E} \left[ T_{i,k}(a_k) + \max_{a_{k+1} \in \mathcal{A}} Q_{k+1}^*(H_{i,k+1}(a_{k+1})) \,\big|\, H_{i,k}(a_k) \right].
\end{equation}
Here, \( T_{i,k}(a_k) \) denotes the potential survival (or visit) time at stage \( k \) under treatment \( a_k \). 
\( H_{i,k}(a_k) \) represents the observed history up to stage \( k \), with the final treatment \( A_{i,k} \) replaced by the hypothetical assignment \( a_k \). \( H_{i,k+1}(a_{k+1}) \) extends this history by adding the stage-\( (k+1) \) covariate and a hypothetical treatment \( a_{k+1} \).
The expectation is taken over the distribution of future covariates and potential outcomes under the full treatment sequence, assuming no censoring.
The oracle (true) optimal treatment decision at stage \( k \) is defined as:
\begin{equation}
d_{i,k}^{\text{true}} = \arg\max_{a_k \in \mathcal{A}} Q_k^*(H_{i,k}(a_k)).
\end{equation}

\subsection{Buckley-James Q-Learning Framework for Optimizing DTRs}
The Buckley–James Q-learning framework integrates the principles of reinforcement learning with robust survival analysis techniques to address challenges associated with right-censored data. This framework estimates the Q-function at each stage by leveraging the Buckley–James imputation method, which accounts for censored survival times, enabling accurate estimation of treatment effects and cumulative rewards. 
For simplicity, in this section, we use \( H_{i,k} \) to represent the covariate and treatment history up to stage \( k \). 
In later sections, within the counterfactual framework, we will use the notation \( H_{i,k}(a_k) \) to explicitly represent the history that would have been observed had treatment \( a_k \) been assigned at stage \( k \).

\subsubsection{Stage-Specific Survival Time Imputation}
In the presence of right-censoring, the observed survival time (or visit time) \( Y_{i,k} \) only partially reflects the true survival time \( T_{i,k} \) for censored observations (\( \delta_{i,k} = 0 \)). To address this limitation, the Buckley-James method \citep{bu79, la91, ji06} imputes censored survival times by computing the conditional expectation:
\begin{equation}
Y_{i,k}^* = \mathbb{E}[T_{i,k} \mid Y_{i,k}, \delta_{i,k}, H_{i,k}],
\end{equation}
where:
\begin{equation}
Y_{i,k}^* = \delta_{i,k}T_{i,k} + (1 - \delta_{i,k}) \mathbb{E}[T_{i,k} \mid T_{i,k} > C_{i}, H_{i,k}].
\end{equation}

For censored observations (\( \delta_{i,k} = 0 \)), the conditional expectation is approximated using the Kaplan-Meier estimator of the residual distribution. Specifically, the imputed survival time at stage \( k \) is expressed as:
\begin{equation} \label{e1}
\hat{Y}_{i,k}(\beta_k) = \delta_{i,k}T_{i,k} + (1 - \delta_{i,k}) \left[ \frac{\int_{e_{i,k}(\beta_k)}^\infty u \, d\hat{F}_k(u)}{1 - \hat{F}_k(e_{i,k}(\beta_k))} + H_{i,k} \beta_k \right].
\end{equation}
Here, \( e_{i,k}(\beta_k) = Y_{i,k} - H_{i,k} \beta_k \) represents the residuals computed for a given regression parameter \( \beta_k \), and \( \hat{F}_k(u) \) is the Kaplan-Meier estimate of the residual distribution.

The Kaplan-Meier estimator \( \hat{F}_k(u) \) is calculated based on the observed residuals and censoring indicators. It is defined as:
\begin{equation}
\hat{F}_k(u) = 1 - \prod_{j: e_{j,k}(\beta_k) \leq u} \left( 1 - \frac{\delta_{j,k}}{\sum_{l=1}^n I\{e_{l,k}(\beta_k) \geq e_{j,k}(\beta_k)\}} \right).
\end{equation}
In this expression, \( \delta_{j,k} \) indicates whether the residual \( e_{j,k}(\beta_k) \) is observed (\( \delta_{j,k} = 1 \)) or censored (\( \delta_{j,k} = 0 \)). The denominator \( \sum_{l=1}^n I\{e_{l,k}(\beta_k) \geq e_{j,k}(\beta_k)\} \) represents the number of residuals at risk at \( e_{j,k}(\beta_k) \).
Once \( \hat{F}_k(u) \) is obtained, the resulting Buckley-James estimator at stage \( k \) is defined as the solution to the estimating equation:
\begin{equation}
\sum_{i=1}^{n} (H_{i,k} - \bar{H}_k) \{\hat{Y}_{i,k}(\beta_k) - H_{i,k} \beta_k\} = 0,
\end{equation}
where \( \bar{H}_k = n^{-1} \sum_{i=1}^{n} H_{i,k} \) represents the mean of the covariates at stage \( k \). However, solving this equation is challenging due to discontinuity and non-monotonicity in \( \beta_k \), which involves the estimation of \( \hat{F}_k(u) \) as well.

To overcome these computational difficulties, a Nelder-Mead simplex algorithm \citep{ji06} was proposed to solve a modified estimating equation iteratively at each stage \( k \). Define the modified least-squares estimating equation as:
\begin{equation}\label{efun2}
U_k(\beta_k,b_k) = \sum_{i=1}^{n} (H_{i,k} - \bar{H}_k) \{\hat{Y}_{i,k}(b_k) - H_{i,k} \beta_k\}.
\end{equation}
Solving the equation \( U_k(\beta_k, b_k) = 0 \) for \( \beta_k \) with \( b_k \) fixed yields the following closed-form solution:
\begin{equation} 
\beta_k = L_k(b_k) = \left\{ \sum_{i=1}^{n} (H_{i,k} - \bar{H}_k)(H_{i,k} - \bar{H}_k)^T \right\}^{-1} 
\left[ \sum_{i=1}^{n} (H_{i,k} - \bar{H}_k) \{\hat{Y}_{i,k}(b_k) - \bar{Y}_k(b_k)\} \right].
\end{equation}
Jin et al. \cite{ji06} show that if the initial estimator \( \tilde\beta_k^{(0)} \) is consistent and asymptotically normal, then the \( l \)th step estimator \( \tilde\beta_k^{(l)} = L_k(\tilde\beta_k^{(l-1)}) \) preserves these asymptotic properties for any \( l \geq 1 \).  
Finally, the Buckley-James estimator at stage \( k \) is then defined as:
\begin{equation}
\tilde\beta_k = \lim_{l\to\infty} L_k(\tilde\beta_k^{(l-1)}).
\end{equation}
The imputed response outcome at stage \( k \), denoted as \( \hat{Y}_{i,k} = \hat{Y}_{i,k} (\tilde\beta_k) \), is then computed using equation \eqref{e1} with the estimated coefficient \( \tilde{\beta}_k \).  
For practical implementation, the initial estimator \( \tilde\beta_k^{(0)} \) is typically set to the least-squares estimator using only uncensored observations at stage \( k \). This approach performs well unless the censoring rate is too high, in which case additional adjustments may be required.

Additionally, the Buckley-James method can be extended to handle double censoring. To address incomplete data, Choi et al. \cite{ch21} proposed replacing the observed \( Y_{i,k} \) with its conditional expectation given the censoring indicators:
\begin{equation}
Y_{i,k}^* = \mathbb{E}[T_{i,k} \mid Y_{i,k}, \delta_{i,k}, H_{i,k}].
\end{equation}
The imputed failure time \( Y_{i,k}^* \) at stage \( k \) is then defined as:
\begin{equation}
Y_{i,k}^* = \delta_{1i,k} T_{i,k} + 
\delta_{2i,k} \mathbb{E}[T_{i,k} \mid T_{i,k} > R_{i,k}, H_{i,k}] +
\delta_{3i,k} \mathbb{E}[T_{i,k} \mid T_{i,k} < L_{i,k}, H_{i,k}].
\end{equation}
where  
\( \delta_{1i,k} \) indicates that \( T_{i,k} \) is fully observed,  
\( \delta_{2i,k} \) indicates right censoring at \( R_{i,k} \),  
\( \delta_{3i,k} \) indicates left censoring at \( L_{i,k} \).
For a detailed discussion on the extension of the Buckley-James method to double censoring, see \cite{ch21, lee2024censored}.

\subsubsection{Recursive Estimation of Q-Functions}

The Q-function \( Q_k(H_{i,k}) \) represents the expected survival reward starting from stage \( k \) under treatment \( A_{i,k} \), given history \( H_{i,k} \). It is estimated recursively, beginning with the final stage \( K \) and moving backward to the first stage.

At the final stage \( K \), the Q-function is initialized using the Buckley–James imputed survival time:
\begin{equation}
\widehat{Q}_K(H_{i,K}) = \hat{Y}_{i,K}(\tilde{\beta}_K),
\end{equation}
where \( \hat{Y}_{i,K}(\tilde{\beta}_K) \) is the imputed survival time under the observed treatment \( A_{i,K} \), obtained via the Buckley–James method.
For earlier stages (\( k = K-1, \ldots, 1 \)), pseudo-outcomes are constructed to combine the immediate imputed survival time with the maximal future survival reward based on estimated Q-functions. Specifically, the pseudo-outcome is defined as:
\begin{equation}
\tilde{Y}_{i,k} = \hat{Y}_{i,k}(\tilde{\beta}_k) + \max_{a_{k+1}} \widehat{Q}_{k+1}(H_{i,k+1}),
\end{equation}
where \( \hat{Y}_{i,k}(\tilde{\beta}_k) \) is the Buckley–James imputed survival time at stage \( k \), and \( \widehat{Q}_{k+1}(H_{i,k+1}) \) is the estimated Q-function from stage \( k+1 \).
At each stage \( k \), the Q-function is estimated by fitting a linear regression model to the pseudo-outcomes:
\begin{equation}
\widehat{\beta}_k = \arg\min_{\beta_k} \sum_{i=1}^n \left( \tilde{Y}_{i,k} - H_{i,k} \beta_k \right)^2.
\end{equation}
The resulting stage-\( k \) Q-function estimate is:
\begin{equation}
\label{eq:Qhat}
\widehat{Q}_k(H_{i,k}) = H_{i,k} \widehat{\beta}_k.
\end{equation}
This recursive procedure uses future-stage Q-estimates as part of the target outcome for current-stage learning, enabling backward induction for treatment optimization. 
Importantly, the use of Buckley–James imputation at each stage ensures that right-censoring is appropriately addressed in estimating expected survival outcomes.

\subsubsection{Optimal Treatment Decision}

At each stage \( k = K, \ldots, 1 \), the goal is to select the treatment that maximizes the expected cumulative survival outcome. The estimated optimal treatment rule is defined as
\begin{equation}
\widehat{d}_{i,k} = \arg\max_{a_k} \widehat{Q}_k(H_{i,k}(a_k)),
\end{equation}
where \( H_{i,k}(a_k) \) denotes the same history \( H_{i,k} = (B_{i,0}, X_{i,1}, A_{i,1}, \ldots, X_{i,k}, A_{i,k}) \) but with the final treatment \( A_{i,k} \) replaced by the hypothetical assignment \( a_k \).

For example, suppose two treatment options, A and B, are available. Although only one treatment is observed per individual, we evaluate the Q-function under both hypothetical assignments:
\begin{equation}
\widehat{Q}_k^{(\text{A})} = \widehat{Q}_k(H_{i,k}(\text{A})), \quad 
\widehat{Q}_k^{(\text{B})} = \widehat{Q}_k(H_{i,k}(\text{B})).
\end{equation}
The estimated decision rule then selects the treatment yielding the higher estimated Q-value, as defined in Equation~\eqref{eq:Qhat}:
\begin{equation}
\widehat{d}_{i,k} = 
\begin{cases}
\text{A} & \text{if } \widehat{Q}_k(H_{i,k}(\text{A})) > \widehat{Q}_k(H_{i,k}(\text{B})), \\
\text{B} & \text{otherwise}.
\end{cases}
\end{equation}
Similarly, the true optimal treatment rule is defined as
\begin{equation}
d_{i,k}^{\text{true}} = 
\begin{cases}
\text{A} & \text{if } Q_k^*(H_{i,k}(\text{A})) > Q_k^*(H_{i,k}(\text{B})), \\
\text{B} & \text{otherwise},
\end{cases}
\end{equation}
where the true Q-function is defined in Equation~\eqref{eq:true_q_function}.

Theoretically, if the Buckley-James imputed survival times \( \hat{Y}_{i,k}(\tilde{\beta}_k) \) are consistent estimators of the true survival times \( T_{i,k} \), then the learned policy is expected to converge in probability to the optimal treatment decision:
\begin{equation}
\widehat{d}_{i,k} \xrightarrow{p} d_{i,k}^{\text{true}}, \quad \text{as } n \to \infty.
\end{equation}
However, the validity of this result depends on the behavior of the Kaplan-Meier residuals in equation \eqref{e1}, which introduces additional variance. Ensuring the consistency of \( \hat{Y}_{i,k}(\tilde{\beta}_k) \) may require additional regularity conditions related to the censoring mechanism and the estimation of residuals.
Further theoretical analysis and empirical investigation are necessary to determine whether the imputed values \( \hat{Y}_{i,k} (\tilde{\beta}_k) \) reliably approximate \( T_{i,k} \) as the sample size increases.

\section{Simulation Study}
This study evaluates the performance of the Buckley-James Q-Learning framework in the context of a clinical trial with right-censored survival outcomes. We assess the accuracy of stage-specific survival time imputation and Q-function estimation across various sample sizes (\(n\)), reflecting typical settings in multi-stage randomized clinical trials.

We first focus on simulating right-censored survival data under a one-stage dynamic treatment regime (DTR) framework and later extend the analysis to the two-stage setting.
Patient-level covariates included binary sex, sampled as \(\text{Sex}_i \sim \text{Bernoulli}(0.5)\), and a stage-specific tumor size \(\text{TumorSize}_{i,k} \sim \text{Unif}(-1, 3)\), representing progression relative to baseline.
The true survival time (or visit time) at stage \( k \) for each patient was generated from a linear model that allows for treatment effect modification:
\[
T_{i,k} = \beta_0 + \beta_1 \cdot \text{Sex}_i + \beta_2 \cdot \text{TumorSize}_{i,k} + \beta_3 \cdot \mathbbm{1}(A_{i,k} = 1) + \beta_4 \cdot \text{TumorSize}_{i,k} \cdot \mathbbm{1}(A_{i,k} = 1) + \varepsilon_{i,k},
\]
where \(A_{i,k} \in \{0, 1\}\) denotes the treatment assignment at stage \(k\) (0 for B, 1 for A), randomized with equal probability across individuals and stages, and \(\varepsilon_{i,k} \sim \mathcal{N}(0, 1)\) represents random noise.
The coefficient \(\beta_4 > 0\) introduces positive treatment effect heterogeneity with respect to tumor size, such that patients with greater tumor progression (i.e., larger \(\text{TumorSize}_{i,k}\)) derive greater benefit from Treatment A.
The true model used parameter values: $\beta_0 = 10$, $\beta_1 = 0.1$, $\beta_2 = -1$, $\beta_3 = 0.01$, and $\beta_4 = 1.3$.
Censoring times (\( C_{i} \)) were independently sampled from a uniform distribution defined between the 20th and 80th percentiles of the total survival time (i.e., \( T_{i,1} \) for the single-stage setting and \( T_{i,1} + T_{i,2} \) for the two-stage setting).
The observed time and censoring indicator were defined as:
\[
Y_{i,k} = \min(T_{i,k}, C_{i}), \quad \delta_{i,k} = \mathbbm{1}(T_{i,k} \leq C_{i}).
\]
We implemented Q-value estimation using three approaches: true (oracle) Q-values, Buckley-James (BJ) imputation using linear regression, and Cox proportional hazards modeling.

Imputation for missing covariates was performed using predictive mean matching (PMM) via the \texttt{mice} package. For censored survival outcomes, we employed the Buckley-James (BJ) method under the accelerated failure time (AFT) model. At each stage \( k \in \{1, 2\} \), the BJ estimator replaces censored observations with their conditional expectations given covariates and censoring status:
\begin{equation}
\label{eq:bj-imputation}
\hat{Y}_{i,k}^{\text{BJ}} = \delta_{i,k} T_{i,k} + (1 - \delta_{i,k}) \, \mathbb{E}[T_{i,k} \mid T_{i,k} > C_{i}, H_{i,k}],
\end{equation}
where \( \delta_{i,k} \) is the event indicator, \( C_{i} \) is the censoring time, and \( H_{i,k} = (\text{TumorSize}_{i,k}, \text{Sex}_i) \) denotes the covariates at stage \( k \). The conditional expectation was approximated using the Kaplan–Meier estimator of the residual distribution on the log-time scale. The BJ procedure iteratively updates regression coefficients based on these partially imputed values until convergence.

For comparison, we also implemented a Cox regression–based imputation. At each stage \( k \), we fit a Cox proportional hazards model:
\[
\lambda_k(t \mid H_{i,k}) = \lambda_{0,k}(t) \exp\left( \beta_k^\top H_{i,k} \right),
\]
where \( \lambda_{0,k}(t) \) is the baseline hazard function and \( \beta_k \) the coefficient vector. The expected survival time was approximated by numerical integration of the estimated survival function:
\begin{equation}
\label{eq:cox-imputation}
\hat{Y}_{i,k}^{\text{Cox}} = \int_0^\infty \exp\left( -\hat{\Lambda}_{0,k}(t) \exp(\beta_k^\top H_{i,k}) \right) dt,
\end{equation}
with \( \hat{\Lambda}_{0,k}(t) \) estimated via the Breslow estimator. Unlike the BJ method, the Cox-based approach relies on the proportional hazards assumption and provides a semi-parametric alternative for handling censored data.

We first consider the single-stage case \( k = 1 \), and later extend to \( k \in \{1, 2\} \). At each stage \( k \), the true Q-functions represent the counterfactual expected survival time under each treatment, defined as
\begin{equation}
\label{eq:true-Q-functions}
Q_{i,k}^{(\text{A})} = \mathbb{E}[T_{i,k}(\text{A}) \mid H_{i,k}(\text{A})], \quad
Q_{i,k}^{(\text{B})} = \mathbb{E}[T_{i,k}(\text{B}) \mid H_{i,k}(\text{B})]
\end{equation}
where \( T_{i,k}{(a_k)} \) is the potential outcome under treatment \( a_k \in \{\text{A}, \text{B}\} \), and \( H_{i,k}(a_k) \) refers to the same history $H_{i,k}$ with the final treatment \( A_{i,k} \) replaced by a hypothetical assignment \( a_k \in \{\text{A}, \text{B}\} \).
In our simulation, these counterfactual expectations are generated from the following linear models:
\[
\begin{aligned}
Q_{i,k}^{(\text{A})} &= \beta_0 + \beta_1 \cdot \text{Sex}_i + \beta_2 \cdot \text{TumorSize}_{i,k} + \beta_3 + \beta_4 \cdot \text{TumorSize}_{i,k}, \\
Q_{i,k}^{(\text{B})} &= \beta_0 + \beta_1 \cdot \text{Sex}_i + \beta_2 \cdot \text{TumorSize}_{i,k}.
\end{aligned}
\]
The interaction term \( \beta_4 > 0 \) induces treatment effect heterogeneity, indicating that the benefit of Treatment~A increases with tumor size. 
Note that tumor size is sampled from a $\text{Unif}(-1, 3)$ distribution, representing progression relative to baseline.
Based on observed and imputed survival data, we estimate the Q-functions using linear regression applied to Buckley–James imputed outcomes or Cox-based survival predictions. 
The estimated Q-functions at each stage \( k \in \{1, 2\} \) are defined as:
\[
\begin{aligned}
\widehat{Q}_{i,k}^{(\text{A})} &= \hat{\beta}_0 + \hat{\beta}_1 \cdot \text{Sex}_i + \hat{\beta}_2 \cdot \text{TumorSize}_{i,k} + \hat{\beta}_3 + \hat{\beta}_4 \cdot \text{TumorSize}_{i,k}, \\
\widehat{Q}_{i,k}^{(\text{B})} &= \hat{\beta}_0 + \hat{\beta}_1 \cdot \text{Sex}_i + \hat{\beta}_2 \cdot \text{TumorSize}_{i,k}.
\end{aligned}
\]
These estimated Q-functions reflect a linear model structure that captures treatment effect heterogeneity through interaction with tumor size.
Right censoring is addressed via Buckley–James imputation under the accelerated failure time (AFT) model (Equation~\eqref{eq:bj-imputation}) or via Cox regression–based estimation (Equation~\eqref{eq:cox-imputation}), both of which are applied to construct Q-function estimates used for treatment decision-making.

\begin{figure}[h!]
  \centering
  \begin{subfigure}[b]{0.75\linewidth}
    \centering
    \includegraphics[width=0.7\textwidth]{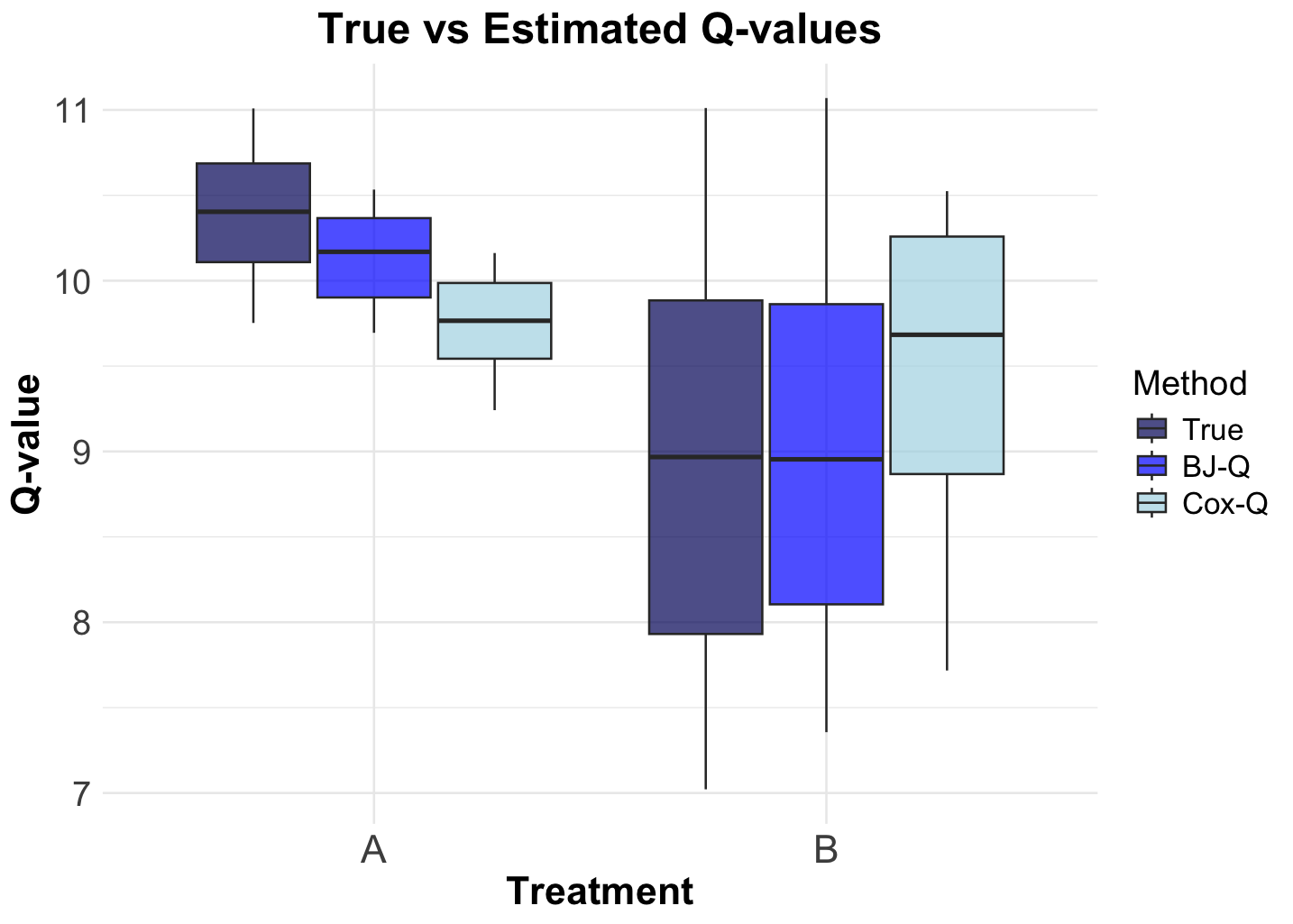}
    \caption{($n$, Missing Rate, Censoring Rate) = (100, 0.5, 0.5)}
  \end{subfigure}
  \vskip\baselineskip
  \begin{subfigure}[b]{0.75\linewidth}
    \centering
    \includegraphics[width=0.7\textwidth]{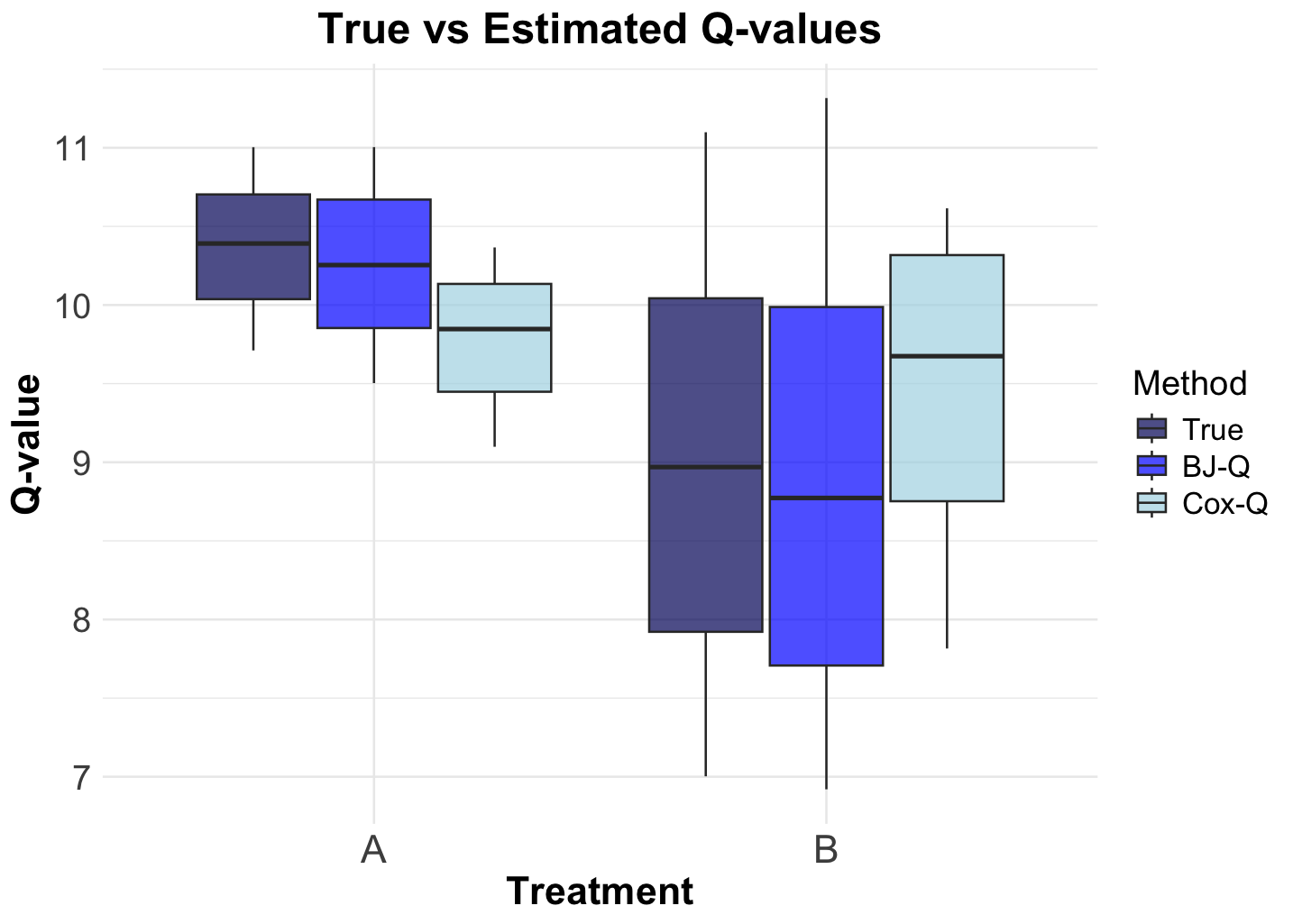}
    \caption{($n$, Missing Rate, Censoring Rate) = (500, 0.5, 0.5)}
  \end{subfigure}
  \vskip\baselineskip
  \begin{subfigure}[b]{0.75\linewidth}
    \centering
    \includegraphics[width=0.7\textwidth]{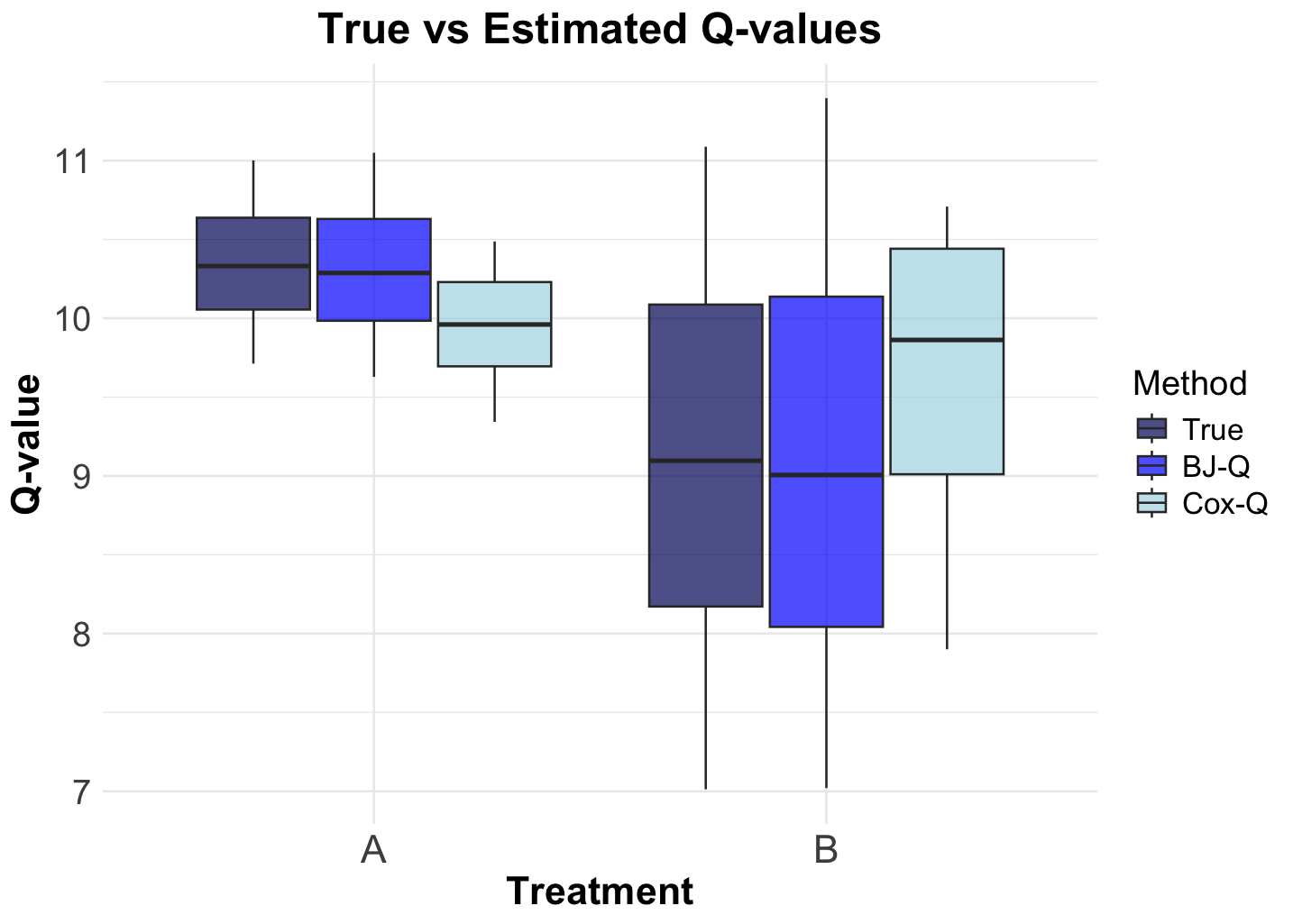}
    \caption{($n$, Missing Rate, Censoring Rate) = (1000, 0.5, 0.5)}
  \end{subfigure}
\caption{Comparison of Estimated Q-values for Treatments A and B using a Single-Stage Dynamic Treatment Regime. Each panel corresponds to a single simulated replicate with 50\% right-censoring. Boxplots compare true Q-values to those imputed using Buckley–James Q-learning (BJ-Q) and Cox-based Q-learning (Cox-Q), across increasing sample sizes.}
  \label{fig:qvalue_comparison_single}
\end{figure}

Figure~\ref{fig:qvalue_comparison_single} displays boxplots of true and estimated Q-values for Treatments A and B under a one-stage dynamic treatment regime across increasing sample sizes ($n = 100$, $500$, and $1000$), with a fixed missingness rate of 50\% and censoring rate of 50\%.
The Buckley–James Q-learning (BJ-Q) method consistently produces Q-value estimates that closely align with the true Q-values across all sample sizes. As $n$ increases, BJ-Q exhibits reduced variability and accurately captures the separation between Treatment A and B. In contrast, the Cox-based Q-learning (Cox-Q) method demonstrates persistent bias, particularly for Treatment A, where it fails to reflect the true treatment effect heterogeneity. Even at $n = 1000$, Cox-Q does not fully recover the true Q-value structure.
These results underscore the advantage of BJ-Q in estimating individual-level treatment effects under right censoring.

To evaluate treatment decision accuracy, we compared estimated optimal treatment assignments against oracle decisions derived from the true Q-functions, defined prior to censoring. The true optimal decision rule assigns Treatment~A if the counterfactual expected survival under Treatment~A exceeds that under Treatment~B:
\[
d_{i,1}^{\text{true}} = \mathbbm{1}\left\{ Q_{i,1}^{(\text{A})} > Q_{i,1}^{(\text{B})} \right\},
\]
where \( Q_{i,1}^{(\text{A})} \) and \( Q_{i,1}^{(\text{B})} \) are the true counterfactual expectations based on the known data-generating mechanism, defined according to Equation~\eqref{eq:true-Q-functions}.
Then, estimated treatment decisions were defined analogously using model-based Q-function estimates:
\[
\begin{aligned}
\widehat{d}_{i,1}^{\text{BJ}} &= \mathbbm{1}\left\{ \widehat{Q}_{i,1}^{{(\text{A})}, \text{BJ}} > \widehat{Q}_{i,1}^{{(\text{B})}, \text{BJ}} \right\}, \\
\widehat{d}_{i,1}^{\text{Cox}} &= \mathbbm{1}\left\{ \widehat{Q}_{i,1}^{{(\text{A})}, \text{Cox}} > \widehat{Q}_{i,1}^{{(\text{B})}, \text{Cox}} \right\},
\end{aligned}
\]
where \( \widehat{Q}_{i,1}^{(a_1), \text{BJ}} \) is using the Buckley–James method under treatment \( a_1 \in \{\text{A}, \text{B}\} \), based on an accelerated failure time model with conditional imputation, and \( \widehat{Q}_{i,1}^{(a), \text{Cox}} \) is the corresponding Cox-based estimate.

Each rule assigns the treatment yielding the higher estimated survival benefit. Decision accuracy was computed as the proportion of individuals whose estimated treatment matched the true (oracle) decision:
\[
\text{Accuracy} = \frac{1}{n} \sum_{i=1}^{n} \mathbbm{1} \left\{ \widehat{d}_{i,1}^{\dagger} = d_{i,1}^{\text{true}} \right\},
\]
where \( \dagger \in \{\text{BJ}, \text{Cox}\} \) indicates the estimation method used.

\begin{table}[ht]
\small
\centering
\caption{Summary Statistics (Minimum, 1st Quartile, Median, Mean, 3rd Quartile, and Maximum) for Treatment Decision Accuracy in Single-Stage DTR Across 50 Simulation Replicates}
\label{tab:accuracy_summary}
\begin{tabular}{llcccccc}
\toprule
$n$ & Method & Min. & 1st Qu. & Median & Mean & 3rd Qu. & Max. \\
\midrule
\multirow{2}{*}{100}  
& BJ-Q  & 0.820 & 0.900 & 0.925 & 0.919 & 0.940 & 1.000 \\
& Cox-Q & 0.480 & 0.653 & 0.715 & 0.721 & 0.795 & 0.930 \\
\midrule
\multirow{2}{*}{500}  
& BJ-Q  & 0.924 & 0.949 & 0.958 & 0.961 & 0.974 & 0.996 \\
& Cox-Q & 0.666 & 0.707 & 0.757 & 0.752 & 0.780 & 0.850 \\
\midrule
\multirow{2}{*}{1000} 
& BJ-Q  & 0.948 & 0.961 & 0.973 & 0.973 & 0.986 & 0.998 \\
& Cox-Q & 0.693 & 0.756 & 0.770 & 0.768 & 0.781 & 0.823 \\
\bottomrule
\end{tabular}
\end{table}

Table~\ref{tab:accuracy_summary} summarizes the decision accuracy of the Buckley–James Q-learning (BJ-Q) and Cox-based Q-learning (Cox-Q) methods across varying sample sizes (\( n = 100, 500, 1000 \)), each replicated 50 times. Across all sample sizes, BJ-Q consistently demonstrated higher accuracy than Cox-Q, with notable improvements in both central tendency and variability.
For \( n = 100 \), BJ-Q achieved a median accuracy of 0.925, compared to 0.715 for Cox-Q. As the sample size increased to \( n = 500 \) and \( n = 1000 \), the accuracy of BJ-Q further improved (median: 0.958 and 0.973), approaching perfect classification in some replications (maximum accuracy reaching 0.996 and 0.998, respectively). In contrast, Cox-Q showed slower gains in accuracy with increasing sample size, with median values of 0.757 and 0.770 at \( n = 500 \) and \( n = 1000 \), respectively.
These results highlight the robustness and superior performance of BJ-Q in learning optimal treatment rules under right-censored data, especially in the presence of treatment effect heterogeneity and increasing data availability.

We now extend our analysis to a two-stage dynamic treatment regime. At each stage \( k \in \{1, 2\} \), patients are assigned either Treatment~A or Treatment~B, resulting in four possible treatment sequences: AA, AB, BA, and BB.
For each treatment sequence \( (a_1, a_2) \in \{\text{A}, \text{B}\}^2 \), we define the true cumulative Q-value as the sum of the counterfactual expected survival times across both stages:
\[
Q_{i,\text{cum}}^{(a_1, a_2)} = Q_{i,1}^{(a_1)} + Q_{i,2}^{(a_2)},
\]
where \( Q_{i,k}^{(a_k)} \) is the counterfactual expected survival time at stage \( k \) under treatment \( a_k \in \{\text{A}, \text{B}\} \), defined according to Equation~\eqref{eq:true-Q-functions}.
These cumulative Q-values serve as the oracle benchmark for evaluating estimation performance.
To accommodate right-censored data, we estimate cumulative Q-values under two modeling approaches. For each individual \( i \) and treatment sequence \( (a_1, a_2) \), the estimated cumulative Q-values are defined as:
\[
\begin{aligned}
\widehat{Q}_{i,\text{cum}}^{(a_1, a_2), \text{BJ}} &= \widehat{Q}_{i,1}^{(a_1), \text{BJ}} + \widehat{Q}_{i,2}^{(a_2), \text{BJ}}, \\
\widehat{Q}_{i,\text{cum}}^{(a_1, a_2), \text{Cox}} &= \widehat{Q}_{i,1}^{(a_1), \text{Cox}} + \widehat{Q}_{i,2}^{(a_2), \text{Cox}},
\end{aligned}
\]
where \( \widehat{Q}_{i,k}^{(a_k), \text{BJ}} \) and \( \widehat{Q}_{i,k}^{(a_k), \text{Cox}} \) are the estimated counterfactual survival times at stage \( k \) under treatment \( a_k \in \{\text{A}, \text{B}\} \), obtained using the Buckley–James and Cox-based Q-learning methods, respectively.
Comparing \( \widehat{Q}_{i,\text{cum}}^{(a_1, a_2), \text{BJ}} \) and \( \widehat{Q}_{i,\text{cum}}^{(a_1, a_2), \text{Cox}} \) to the corresponding true cumulative values \( Q_{i,\text{cum}}^{(a_1, a_2)} \) enables comprehensive evaluation of each method’s ability to recover total survival under multi-stage decision-making in the presence of right-censoring.

\begin{figure}[h!]
  \centering
  \begin{subfigure}[b]{0.75\linewidth}
    \centering
    \includegraphics[width=0.75\textwidth]{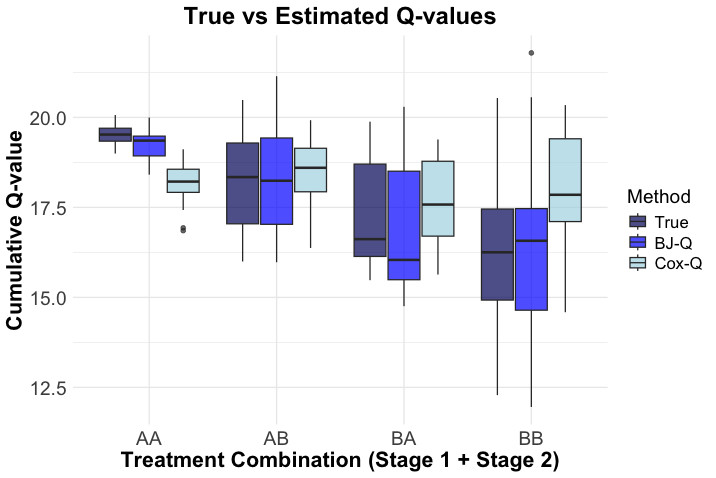}
    \caption{($n$, Missing Rate, Censoring Rate) = (100, 0.5, 0.5)}
  \end{subfigure}
  \vskip\baselineskip
  \begin{subfigure}[b]{0.75\linewidth}
    \centering
    \includegraphics[width=0.75\textwidth]{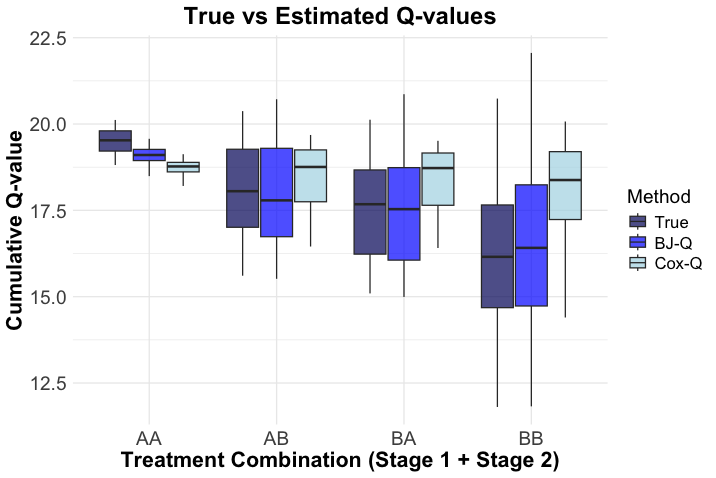}
    \caption{($n$, Missing Rate, Censoring Rate) = (500, 0.5, 0.5)}
  \end{subfigure}
  \vskip\baselineskip
  \begin{subfigure}[b]{0.75\linewidth}
    \centering
    \includegraphics[width=0.75\textwidth]{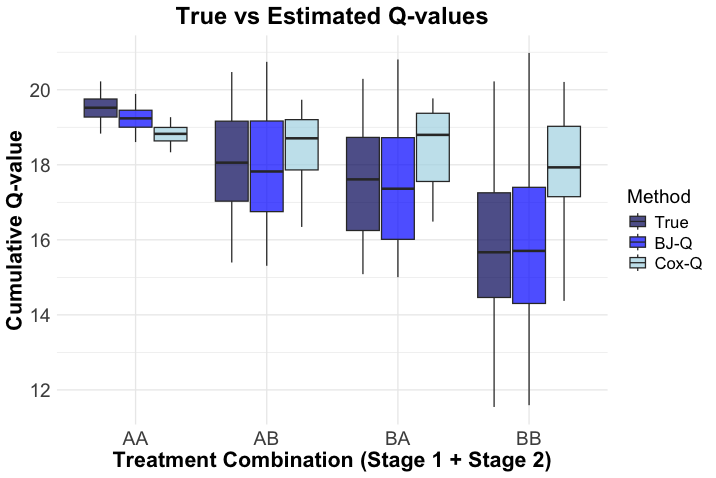}
    \caption{($n$, Missing Rate, Censoring Rate) = (1000, 0.5, 0.5)}
  \end{subfigure}
\caption{Comparison of Cumulative Q-values across Treatment Combinations in a Two-Stage Dynamic Treatment Regime. Each treatment combination (AA, AB, BA, BB) denotes the assigned treatments at stages 1 and 2. Boxplots compare the cumulative true Q-values to those estimated using Buckley–James Q-learning (BJ-Q) and Cox-based Q-learning (Cox-Q) under 50\% right-censoring.}
\label{fig:qvalue_comparison_double}
\end{figure}

Figure~\ref{fig:qvalue_comparison_double} presents the comparison of cumulative Q-values across treatment combinations in a two-stage dynamic treatment regime (DTR). Each treatment combination (AA, AB, BA, BB) denotes the pair of treatments received at stages 1 and 2, respectively. The cumulative Q-value represents the sum of stage-specific expected outcomes, reflecting the total survival benefit under a given treatment sequence.
The boxplots display the distribution of cumulative Q-values for each combination, comparing the oracle (true) values against those estimated using Buckley–James Q-learning (BJ-Q) and Cox-based Q-learning (Cox-Q). The data were simulated under 50\% right-censoring, and imputation was conducted accordingly for censored observations. 
As illustrated, BJ-Q yields Q-value distributions that align more closely with the true values across all treatment sequences.

We evaluated the accuracy of estimated treatment decisions under a two-stage dynamic treatment regime using the Q-learning framework, which determines the optimal treatment sequence. 
The estimated optimal treatment sequence for individual \( i \) was obtained by maximizing the sum of the estimated Q-values over all possible treatment combinations:
\[
(\widehat{d}_{i,1}^{\dagger}, \widehat{d}_{i,2}^{\dagger}) = \arg\max_{(a_1, a_2) \in \{\text{A}, \text{B}\}^2}
\left\{ \widehat{Q}_{i,1}^{(a_1), \dagger} + \widehat{Q}_{i,2}^{(a_2), \dagger} \right\},
\]
where \( \dagger \in \{\text{BJ}, \text{Cox}\} \).
For comparison, the true optimal treatment sequence was defined using oracle Q-functions based on the uncensored counterfactual survival times:
\[
(d_{i,1}^{\text{true}}, d_{i,2}^{\text{true}}) = \arg\max_{(a_1, a_2) \in \{\text{A}, \text{B}\}^2}
\left\{ Q_{i,1}^{(a_1)} + Q_{i,2}^{(a_2)} \right\},
\]
where \( Q_{i,k}^{(a_k)} \) denotes the true counterfactual expected survival time at stage \( k \), under treatment \( a_k \in \{\text{A}, \text{B}\} \), computed from the known data-generating mechanism prior to censoring.
Finally, decision accuracy was defined as the proportion of individuals for whom the estimated treatment sequence matched the true (oracle) sequence across both stages:
\[
\text{Accuracy} = \frac{1}{n} \sum_{i=1}^{n} \mathbbm{1} \left\{
(\widehat{d}_{i,1}^{\dagger}, \widehat{d}_{i,2}^{\dagger}) = (d_{i,1}^{\text{true}}, d_{i,2}^{\text{true}})
\right\},
\]
where \( \dagger \in \{\text{BJ}, \text{Cox}\} \).

\begin{table}[ht]
\small
\centering
\caption{Summary Statistics (Minimum, 1st Quartile, Median, Mean, 3rd Quartile, and Maximum) for Treatment Decision Accuracy at Each Stage and Cumulative Sequence in a Two-Stage DTR Across 50 Simulation Replicates}
\label{tab:accuracy_two_stage}
\begin{tabular}{llcccccc}
\toprule
$n$ & Method & Min. & 1st Qu. & Median & Mean & 3rd Qu. & Max. \\
\midrule
\multirow{6}{*}{100}
& BJ-Q: Stage 1      & 0.810 & 0.910 & 0.945 & 0.934 & 0.960 & 0.990 \\
& BJ-Q: Stage 2      & 0.820 & 0.880 & 0.925 & 0.907 & 0.940 & 0.970 \\
& BJ-Q: Cumulative   & 0.730 & 0.815 & 0.865 & 0.850 & 0.890 & 0.960 \\
& Cox-Q: Stage 1     & 0.550 & 0.692 & 0.760 & 0.737 & 0.787 & 0.920 \\
& Cox-Q: Stage 2     & 0.490 & 0.590 & 0.685 & 0.677 & 0.747 & 0.920 \\
& Cox-Q: Cumulative  & 0.370 & 0.443 & 0.475 & 0.503 & 0.558 & 0.740 \\
\midrule
\multirow{6}{*}{500}
& BJ-Q: Stage 1      & 0.880 & 0.936 & 0.952 & 0.950 & 0.966 & 0.998 \\
& BJ-Q: Stage 2      & 0.852 & 0.905 & 0.941 & 0.933 & 0.961 & 0.974 \\
& BJ-Q: Cumulative   & 0.782 & 0.848 & 0.893 & 0.886 & 0.924 & 0.956 \\
& Cox-Q: Stage 1     & 0.580 & 0.669 & 0.710 & 0.706 & 0.747 & 0.794 \\
& Cox-Q: Stage 2     & 0.596 & 0.694 & 0.711 & 0.719 & 0.753 & 0.848 \\
& Cox-Q: Cumulative  & 0.394 & 0.479 & 0.504 & 0.508 & 0.535 & 0.602 \\
\midrule
\multirow{6}{*}{1000}
& BJ-Q: Stage 1      & 0.921 & 0.942 & 0.952 & 0.956 & 0.974 & 0.992 \\
& BJ-Q: Stage 2      & 0.912 & 0.925 & 0.940 & 0.940 & 0.952 & 0.989 \\
& BJ-Q: Cumulative   & 0.860 & 0.879 & 0.899 & 0.900 & 0.919 & 0.951 \\
& Cox-Q: Stage 1     & 0.649 & 0.688 & 0.707 & 0.712 & 0.731 & 0.795 \\
& Cox-Q: Stage 2     & 0.645 & 0.687 & 0.715 & 0.709 & 0.729 & 0.757 \\
& Cox-Q: Cumulative  & 0.452 & 0.485 & 0.508 & 0.506 & 0.530 & 0.565 \\
\bottomrule
\end{tabular}
\end{table}

Table~\ref{tab:accuracy_two_stage} summarizes the accuracy of treatment decisions across stages~1, 2, and their cumulative sequence in a two-stage dynamic treatment regime, evaluated over 50 simulation replicates for sample sizes \( n = 100, 500, 1000 \). 
For all sample sizes, Buckley-James Q-learning (BJ-Q) consistently achieved higher decision accuracy than Cox-based Q-learning (Cox-Q), across both stages and in cumulative assignment. The mean accuracy for BJ-Q at Stage~1 increased from 0.934 (when \( n = 100 \)) to 0.956 (when \( n = 1000 \)), indicating strong stage-wise estimation even in smaller samples. A similar trend was observed at Stage~2, with BJ-Q mean accuracy improving from 0.907 to 0.940 as \( n \) increased. 
In contrast, Cox-Q yielded lower accuracy across all scenarios. For instance, the mean cumulative accuracy for Cox-Q was 0.503 at \( n = 100 \), improving only slightly to 0.506 at \( n = 1000 \), highlighting its limited precision in capturing optimal treatment sequences under right-censoring. BJ-Q cumulative accuracy remained robust, rising from 0.850 to 0.900, affirming the advantage of BJ-based imputation in dynamic treatment learning under censoring.

Although commonly used, the Cox-Q model demonstrates limitations in accurately recovering total survival under censoring and missing data, likely due to its reliance on the proportional hazards assumption and indirect modeling of survival time. In contrast, the BJ-Q method shows greater robustness, particularly in complex longitudinal settings. As the number of decision stages \( K \) increases in multi-stage dynamic treatment regimes, cumulative bias from imprecise survival estimation can accumulate, underscoring the value of accurate imputation. The BJ-Q approach, which directly models conditional survival times without imposing hazard-based assumptions, is especially well-suited for stage-specific decision-making contexts.

\section{Real Data Analysis}
We applied our proposed methodology to the analysis of the ACTG175 dataset, which is available through the \texttt{speff2trial} R package \citep{juraska2022package}. 
This dataset comprises data from 2,139 HIV-infected patients who were randomly assigned to one of four treatment arms: AZT monotherapy, combination therapy with AZT and didanosine (ddI), combination therapy with AZT and zalcitabine (ddC), or didanosine (ddI) monotherapy.
The principal aim of the trial was to assess the effectiveness of monotherapy versus combination therapy in patients with CD4-T cell counts ranging from 200 to 500/mm \citep{hammer1996trial}.
The variable \texttt{days} represents the time until a significant clinical event, such as a decline in CD4 T cell count by at least 50, progression to AIDS, or death. 
The \texttt{cens} variable serves as a censoring indicator, with \texttt{cens} = 1 indicating observed events and \texttt{cens} = 0 representing censored data. 
Notably, the dataset exhibits a severe censoring rate of approximately 75 percent, and there are also some missing values in the explanatory variables, adding complexity to the data analysis. 
This high rate of censoring and the presence of missing data are particularly challenging and require robust statistical methods capable of effectively handling such incomplete data.
The detailed explanation of the explanatory variables and their impacts on the analysis is further elaborated in \cite{juraska2022package}.

Previous studies have shown that patients previously treated with AZT tend to achieve better outcomes when switched to ddI alone or in combination with AZT, rather than continuing AZT monotherapy \cite{hammer1996trial}. 
Our analysis aims to identify the optimal treatment regimen by comparing the efficacy of switching to ddI monotherapy (\(A_i = 0\)) versus continuing combination therapy with AZT and ddI (\(A_i = 1\)). 
Using the Counterfactual Buckley-James (BJ) Q-learning framework under an accelerated failure time model, we first estimated the optimal treatment assignment by maximizing the expected counterfactual survival. 
Based on this assignment, we then imputed censored survival times.

\begin{figure}[h!]
  \centering 
  \includegraphics[scale=0.26]{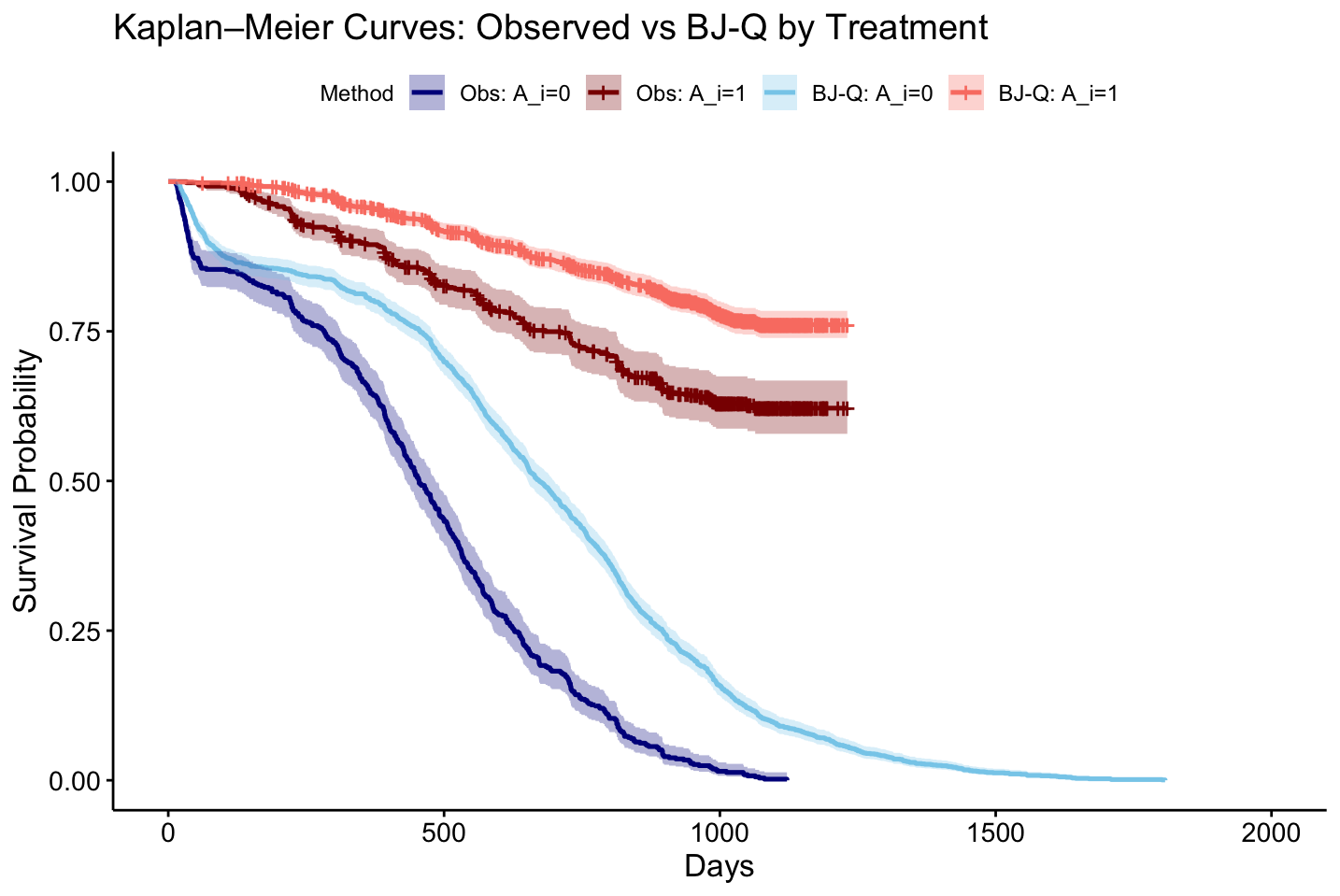}
    \caption{Kaplan-Meier survival curves comparing observed and Buckley-James (BJ-Q) imputed outcomes by treatment group. Curves are stratified by treatment assignment ($A_i = 0$ or $A_i = 1$). Observed curves (dark blue and dark red) reflect right-censored survival data, while BJ-Q curves (light blue and light red) represent imputed survival times under the Buckley-James accelerated failure time model. BJ-Q curves consistently lie above the observed curves, reflecting the correction for censoring.}
  \label{fig:real}
\end{figure}

As shown in Figure~\ref{fig:real}, Kaplan-Meier curves are plotted separately for observed and BJ-imputed survival times by treatment group. 
The dark blue and dark red lines represent the observed survival curves for ddI monotherapy (\(A_i = 0\)) and combination therapy (\(A_i = 1\)), respectively. 
In contrast, the light blue and red curves represent the BJ-imputed survival times under each treatment. 
The red curve (\(A_i = 1\)) consistently dominates the light blue curve (\(A_i = 0\)) across all time points, indicating superior survival outcomes for the combination therapy arm. 
The imputed curves also exhibit tighter confidence bands, reflecting reduced uncertainty compared to the observed data. 
Together, these findings underscore the ability of BJ-Q to uncover latent survival differences obscured by censoring, and provide strong support for combination therapy as the optimal treatment strategy.

It is important to note that the dataset provides only overall survival time after second-stage treatment, without stage-specific information, making full multi-stage Q-learning recursion infeasible. 
To address this, we apply a single-stage version of our BJ-Q learning framework, treating aggregated survival as the outcome and handling censoring via Buckley--James imputation. While designed for multi-stage decision-making, the framework's core principle of optimizing treatment based on counterfactual outcomes remains valid in this simplified setting, and we anticipate that future data with stage-specific details will enable its full application.

\section{Discussion}

This study demonstrates that integrating Q-Learning with an imputed survival reward framework offers a principled and practical approach for handling incomplete data and censoring in healthcare analytics. The proposed Buckley-James Q-Learning (BJ-Q) method accommodates both right and double censoring, enabling robust policy learning in longitudinal decision-making settings. When survival outcomes are consistently observed at each stage, the method can scale to larger values of \(K\), owing to the computational simplicity of the Buckley-James estimator. 
Based on empirical guidance from our simulation study and Jin et al.~\cite{ji06}, we recommend a minimum of 50 uncensored observations per stage to achieve stable estimation.

The current implementation uses the Kaplan–Meier (K–M) estimator to approximate conditional expectations of censored residuals. However, as discussed by Cui et al.~\cite{cui2022consistency}, the K–M estimator may be biased when censoring is informative. To address this limitation, future work will explore alternative approaches such as random survival forests or other nonparametric machine learning methods that can provide more reliable estimates under dependent censoring.

Future research will also focus on extending the BJ-Q framework to high-dimensional covariate spaces where the number of predictors exceeds the sample size. Building on recent developments in penalized methods for censored data~\cite{lee2024censored}, we aim to incorporate variable selection via regularization into the BJ-Q algorithm. To ensure scalability, we will leverage computational strategies such as coordinate descent optimization and parallel computing, making the approach feasible for applications in precision medicine, genomics, and other domains involving complex survival data.

\section*{Acknowledgement(s)}
We thank the two anonymous referees, the Associate Editor, and the Editor-in-Chief for their constructive and helpful suggestions, which have led to substantial improvements in the revised version.

For the sake of transparency and reproducibility, the code and data used for this study can be found in the following GitHub
repository: \url{https://github.com/Jeongjin95/BJ-Q-Learning}.

\section*{Disclosure statement}
No potential conflict of interest was reported by the author(s).

\section*{Funding}
This research received no external funding.

\bibliographystyle{tfq}
\bibliography{interacttfqsample}

\end{document}